# A new photolithography based technique to mass produce microlens+fibre based integralfield units (IFUs) for 2D spectroscopy

Sabyasachi Chattopadhyay[a,*], Vishal Joshi[b], A. N. Ramaprakash[a], Deepa Modi[a], Abhay Kohak[a], and Haeun Chung[e,f]

[a]Inter-University Centre for Astronomy and Astrophysics, Pune, India
[b]Physical Research Laboratory, Ahmedabad, India
[e]Korea Institute for Advanced Study, 85 Heogiro, Dongdaemun-gu, Seoul 02455, Korea
[f]Department of Physics and Astronomy, Seoul National University, 1 Gwanak-ro, Gwanak-gu, Seoul 08826, Korea

**Abstract**

Long-slit astronomical spectroscopy has various limitations when dealing with optimum slit width, atmospheric dispersion, extended source spectroscopy, etc. to name a few. Most of these issues can be solved by the use of optical fibers as the light carrier from the telescope focal plane to the spectrograph. The approach is technically and scientifically flexible in terms of instrument modularity and target acquisition. Implementation of Integral Field Unit (IFU) provides a continuous sampling of extended objects and has a distinct advantage over the single fiber. Using a microlens array in front of the fibers improves the sky coverage by increasing the fill factor. Devasthal Optical Telescope Integral Field Spectrograph (DOTIFS) is a novel instrument being built by the Inter-University Centre for Astronomy & Astrophysics, Pune for the 3.6m Devasthal Optical Telescope (DOT) constructed by Aryabhatta Research Institute of Observational Sciences, Nainital. Each of the 16 DOTIFS IFUs consist of 12×12 spatial elements (spaxels) distributed in a hexagonal honeycomb structure covering 8.7"×7.8" in the sky. Each IFU is made by a photolithography technique to transfer the corresponding microlens array pattern to create a mask which holds the fibers at the focal plane end of an integral field unit. These masks are aligned with the microlens array and fibers are inserted before gluing and polishing. The fiber array can be positioned with a peak positioning error less than 5 $\mu$m from the desired position within a fiber array, compared to a requirement of 10 $\mu$m. The slit end is made by wire EDM cutting technology and fibers are placed with an accuracy of ∼0.3 pixels compared to a 6.75 pixel center-to-center gap between two spectra on the detector. In this paper we provide details of deriving requirements and error budgets. The process of photolithography and the use of generated masks to create an IFU are also discussed. The technique allows very cost effective mass production of IFUs which are very accurately matched with the corresponding microlens array.

## 1 INTRODUCTION

Astronomical spectroscopy is a technique which helps in understanding astrophysical phenomena, their source and the physical processes involved in them. The traditional method of long-slit spectroscopy has various limitations in the form of optimum slit width, atmospheric dispersion, crowded field spectroscopy, etc. to name a few. Most of these issues were solved by the introduction of optical fibers as the light carrier from focal plane to spectrograph (Hill et al., 1980 [3]). The approach is technically and scientifically flexible with respect to instrument modularity and target acquisition. Deployment of the fiber bundle, also called an integral field unit (IFU), provides a continuous sampling of extended objects and has a distinct advantage over the single fiber ([5]). Hence, the method is called area spectroscopy or 2D spectroscopy. The data generated by the IFU forms a 3D data cube of which two axes are spatial dimensions while the third axis is for wavelength. However, an IFU which is only made of fiber, leads to a loss of light from the gap between the fibers. Using a microlens array in front of the fibers improves the sky coverage by increasing the fill factor ([4]). An IFU based on a microlens array and fibers can provide a contiguous sampling of an extended object. These IFUs serve several advantages but suffers from low throughput at the fiber junctions. Another solution to the issues of long-slit spectroscopy is the use of a slicer based IFU ([1]). A slicer transfers the light from the focal plane to the spectrograph via multiple reflecting elements. Slicer based IFUs have a high throughput, but their application is limited to infrared (IR) spectroscopy due to strong scattering in optical wavelengths.

Devasthal Optical Telescope Integral Field Spectrograph (DOTIFS) is a novel instrument for the 3.6m Devasthal Optical Telescope (DOT). The instrument is built for realizing the different science goals around the nearby universe. After various considerations, we have chosen a microlens fiber-based approach for the IFU development for DOTIFS. Several of the current IFU instruments have issues with the accuracy of fiber positioning. We have made a prototype of the microlens fiber based IFU for DOTIFS with photolithography. The technique provides unprecedented accuracy in the positioning of fibers and in the possibility of mass production in an extremely cost-effective manner. In section 3, the method of photolithography will be discussed in detail.

# 2 DESIGN OF INTEGRAL FIELD UNIT

## 2.1 Selection of Microlens Array

We have chosen a regular hexagonal spaxel shape. The spaxel shape is primarily driven by the need to achieve a high fill factor with a minimum scattering of light. Square-shaped microlenses provide better fill factor, but requires a fiber core larger than the same for hexagonal shaped lenslet. Circular lenslet fails to provide the desired fill factor. The microlens array (MLA) forms a hexagonal honeycomb structure of 12×12 spaxels. The on sky size of an IFU is designed to be 8.7”×7.4” for a spatial sampling of 0.8”/300 $\mu$m [2]. We have procured 30 microlens arrays (MLA) with a thickness of 2.32 mm out of which 16 will be used. Each microlens curved side will receive a f/21.486 beam from the magnifier optics assembly sitting between IFU assembly and the Cassegrain sideport selection mirror of the telescope. The magnifier assembly, not only changes focal plane position of the telescope Cassegrain side port but also modifies the image scale to meet the spatial sampling requirements. The 8’ on sky field of view corresponds to 180mm in physical dimension. The microlens converts the incoming beam from the magnifier to f/4.5 beam and creates a spot of 76 $\mu$m diameter at its flat back surface. However, the practically measured diameter of the spot is 80 $\mu$m $\pm 3\mu$m.

## 2.2 Selection of Fiber

We have chosen Optran Polyimide fiber from CeramOptec® as the light carrier element of the IFU. The fiber has a numerical aperture of 0.12 in air. The fiber core needs to cover the beam produced by the MLA at its flat back-plane. The fiber has more than 99% transmission efficiency over 300-800 nm which covers DOTIFS wavelength range. The chosen model has a 100$\mu$m diameter silica core. With a fluorine doped silica cladding, it becomes 110$\mu$m. The fiber also has a protective polyimide jacket which makes the overall diameter 125$\mu$m. The fiber needs to have a positioning accuracy of $\pm 7.5$ $\mu$m for its core to completely engulf the spot created by the microlens.

## 2.3 Characterization of Microlens Array

The position of the spot created by a microlens at its flat side should theoretically be at the crossing of the three diagonals of the hexagon. In practice, the spots are found to have positioning error of several $\mu$m, due to manufacturing defects. Placing the fibers at the ideal spot grid positions will result in loss of light due to an improper matching of the spot and the fiber positions. Thus it is important to know the location of these spots before placing the fibers. The setup shown in Figure 1 is used to characterize the MLAs. A collimated beam is used to illuminate the curved surface of the MLA. Each microlens produces a spot at its flat surface. The flat surface of the MLA was imaged, and the centroid position of each spot is measured. A high signal to noise ratio image provides a sub-pixel centroid position measurement accuracy. The same test is performed by slightly changing the position of the MLA on the detector to ensure the repeatability of the measurement . The difference between any two measurements would provide the measurement error, provided the input beam is uniformly illuminated. Figure 2 shows the spot pattern of an MLA on the CCD. Figure 3 shows the difference between the centroid positions in two orthogonal directions in two measurements. It is found that the the RMS error is $\sim 4$ $\mu$m in two orthogonal direction. Figure 4 shows the mean of the measurement errors of 16 MLAs while the error bars provide the RMS measurement error for the associated MLA.

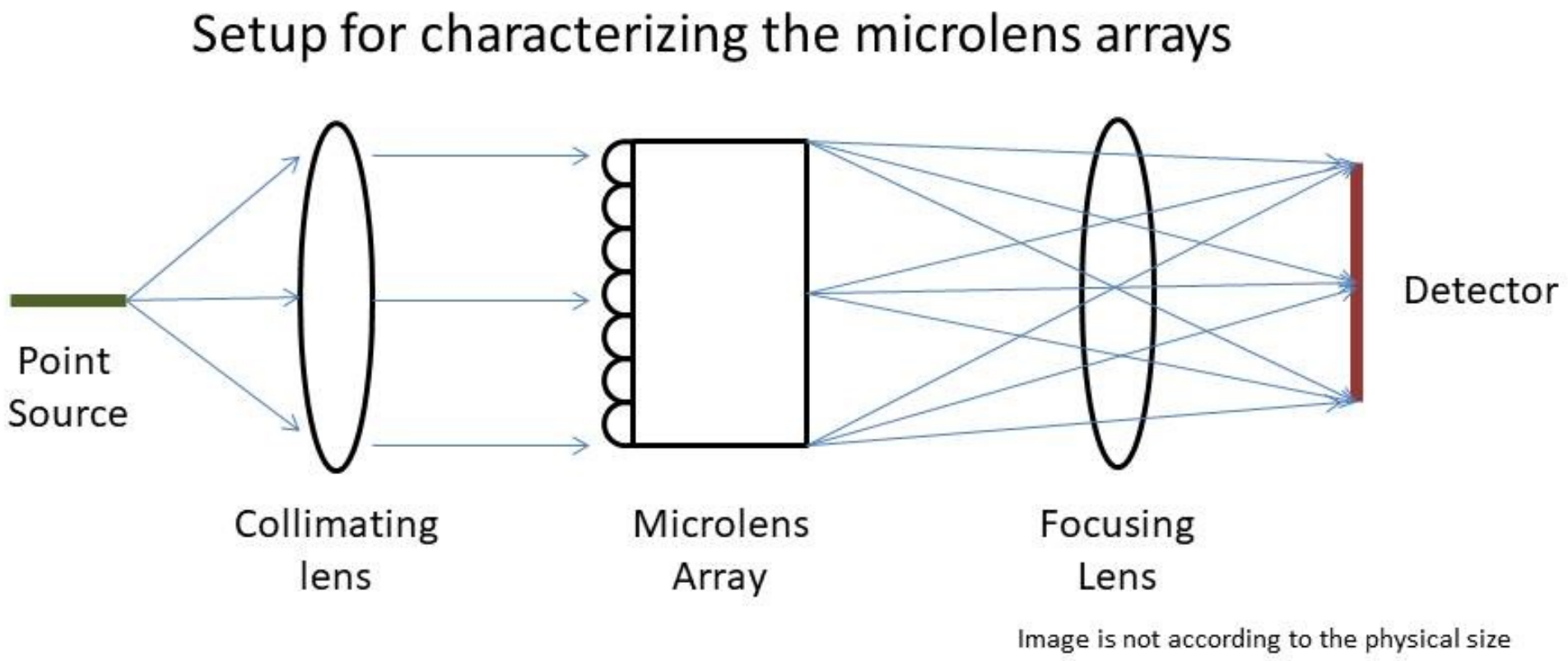

Figure 1: Setup used for characterization of Microlens array.

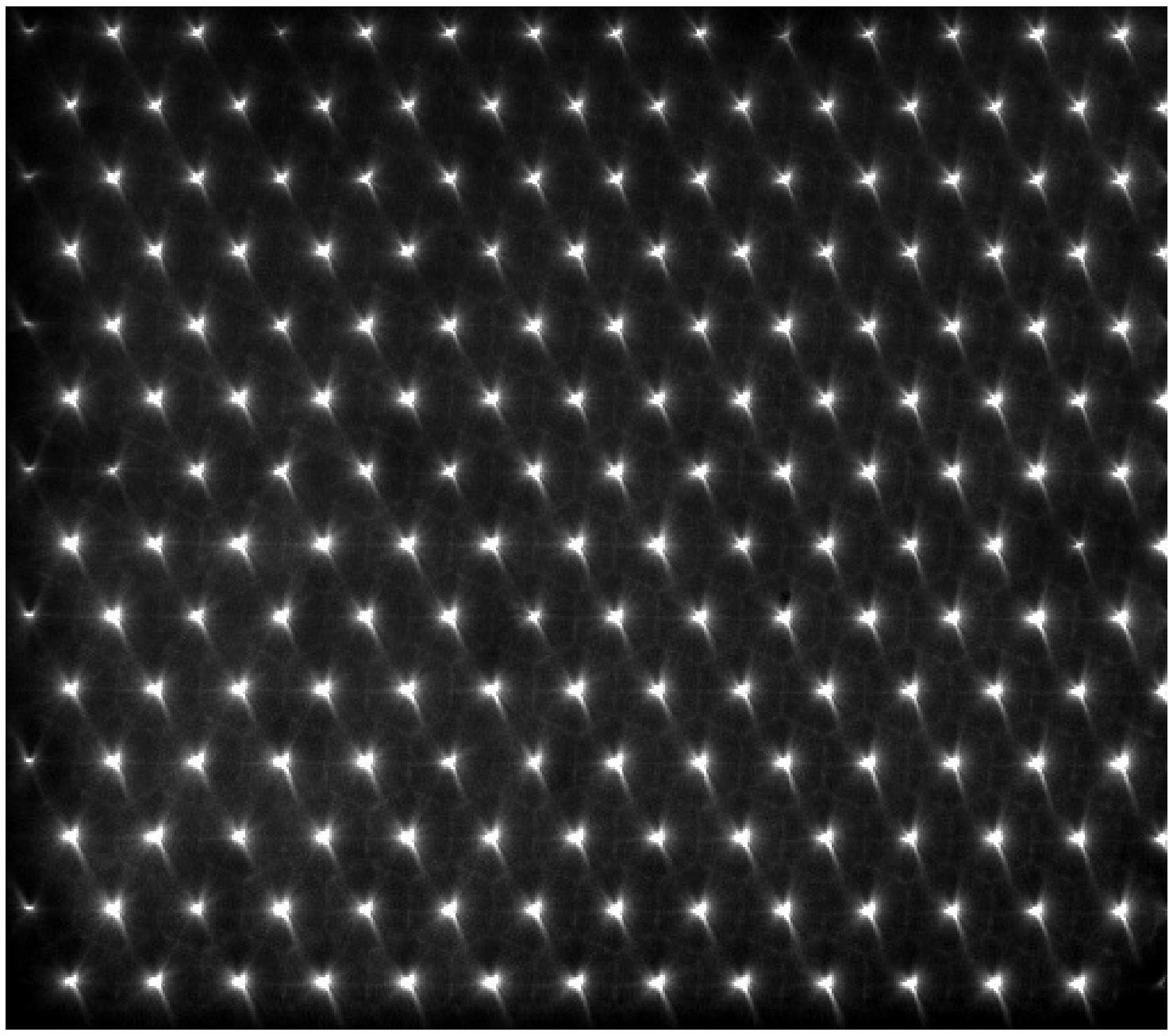

Figure 2: Image of a microlens array flat surface after illuminating the curved surface by collimated light.

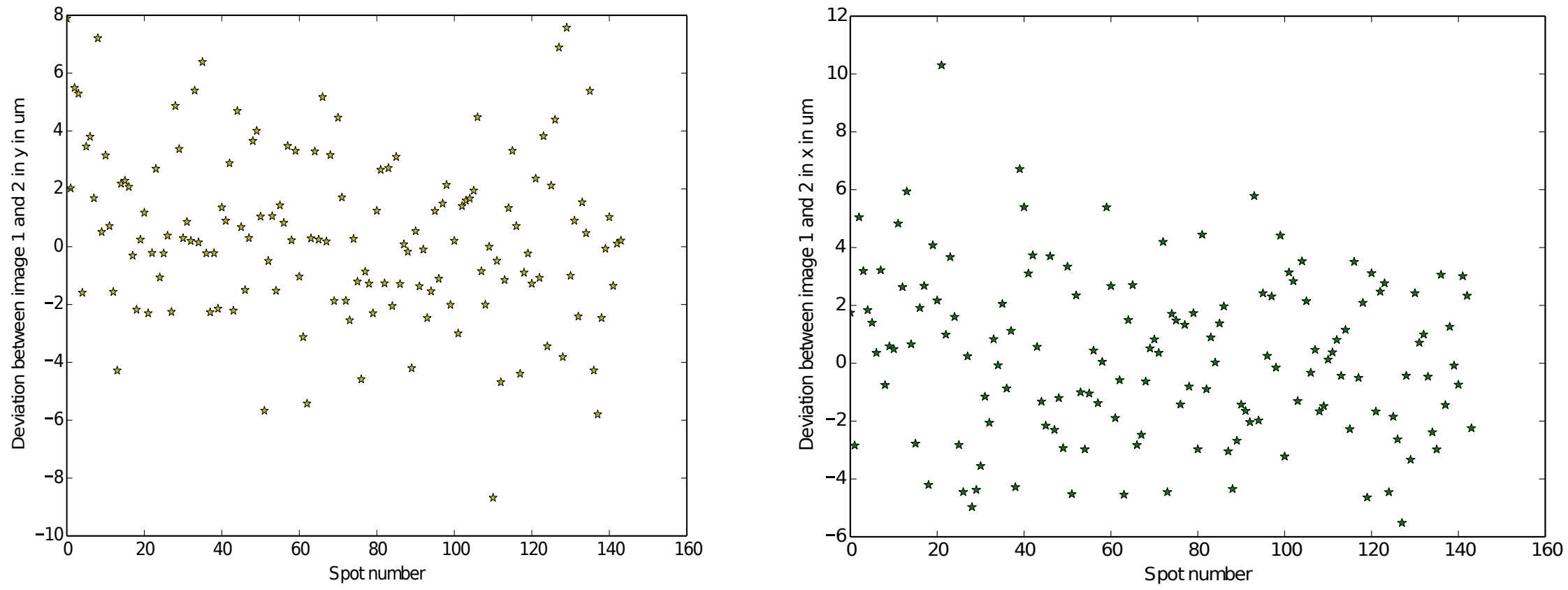

Figure 3: Deviation between position measurement of spots created by a microlens array on its flat surface. The left and right panel shows measurement deviation in two orthogonal direction.

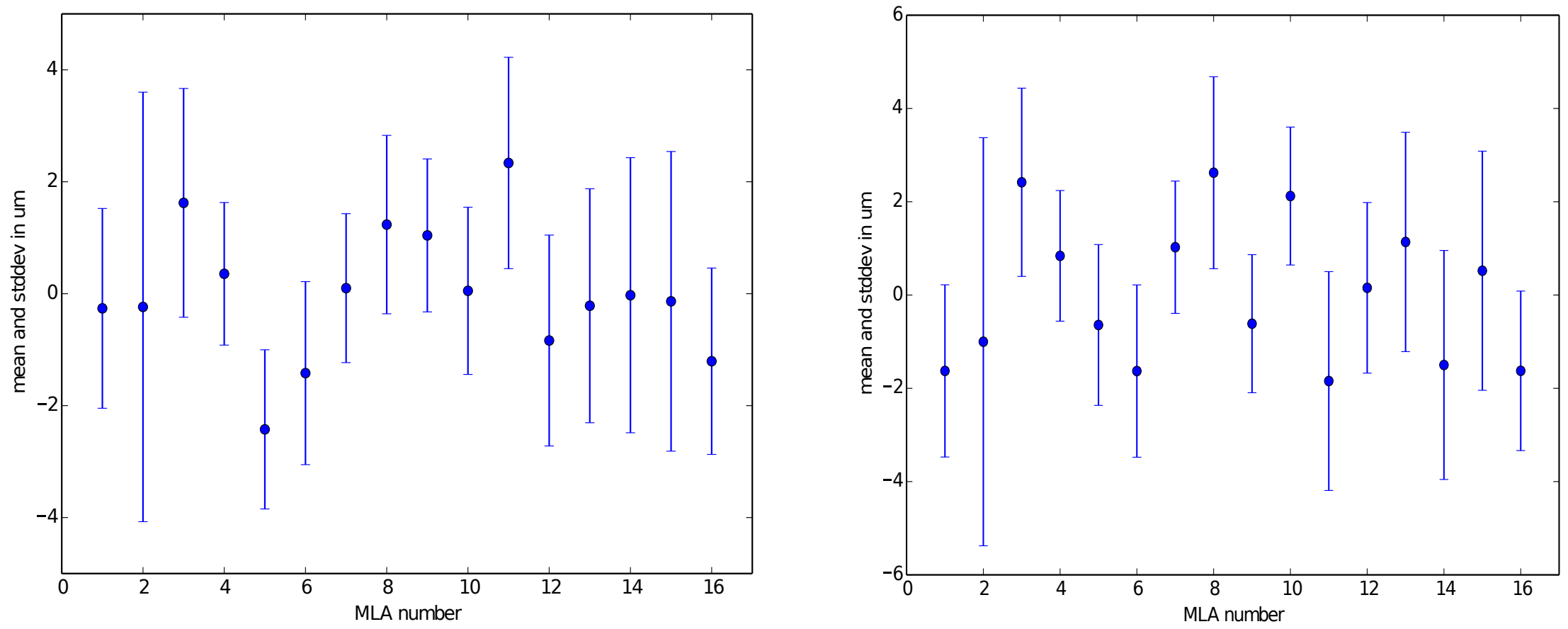

Figure 4: Mean of the measurement errors of spot positions for all 16 microlens arrays. The left and right panel shows measurement deviation in two orthogonal directions. The error bars show the RMS measurement error associated with an MLA.

## 2.4 Fiber Positioning Requirements

In each IFU, 144 fibers need to be positioned behind the MLA. It is challenging to place and glue each fiber separately. Thus it is essential to use a fiber holder (mask) to place the fibers in the desired position. The masks are designed to have holes for each fiber. So the position and the dimensions of the holes in the masks are crucial. The hole diameter can be 7.5 $\mu$m higher than the combined diameter of the fiber with a jacket. The positions of these holes should also be determined by the MLA spot locations. The thickness of the mask is determined from the acceptable tilt of the fiber which is determined to be 0.2°. This tilt angle coupled with an excess hole diameter of 7.5 $\mu$m defines the thickness of the mask to be at least >1.3 mm. We have explored two options to create the mask: laser drilling and photolithography.

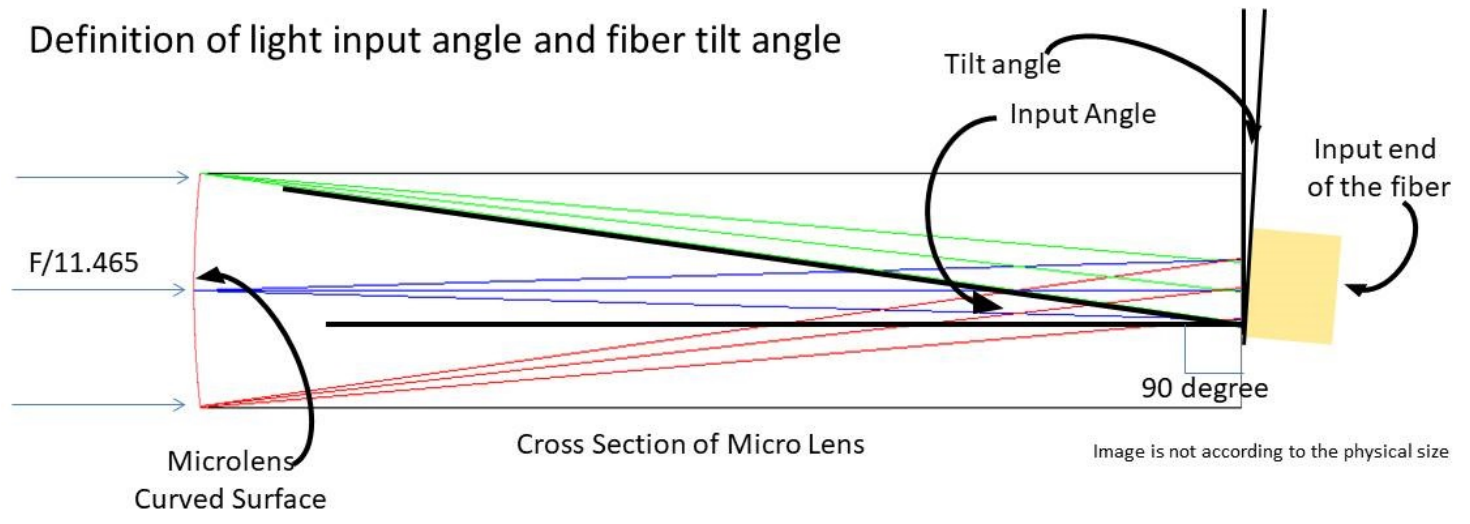

Figure 5: Illustration of the input angle of light into the fiber core, and the tilt angle between Microlens flat surface against fiber input face.

## 2.5 Design of the Slit-end

DOTIFS has eight identical spectrograph units. Each spectrograph is fed by the fibers from two IFUs and forms a series of spectra on one CCD of 2k×4k size, where the dispersion is along the wider axis. A total of 288 spectra are fitted on to 2048 pixels (each pixel is 15 $\mu$m square), with 2.5 pixel FWHM per spectrum and 6.75 pixels center-to-center gap as shown in Figure 6. The demagnification of the spectrograph optics is roughly ~0.384. The recreated slit for the spectrograph should be 80 mm in length and should contain fibers from two IFUs as shown in Figure 7. The center-to-center gap between two consecutive fibers is 270 $\mu$m, with a spacing of 700 $\mu$m kept between the two IFUs to mark the division.

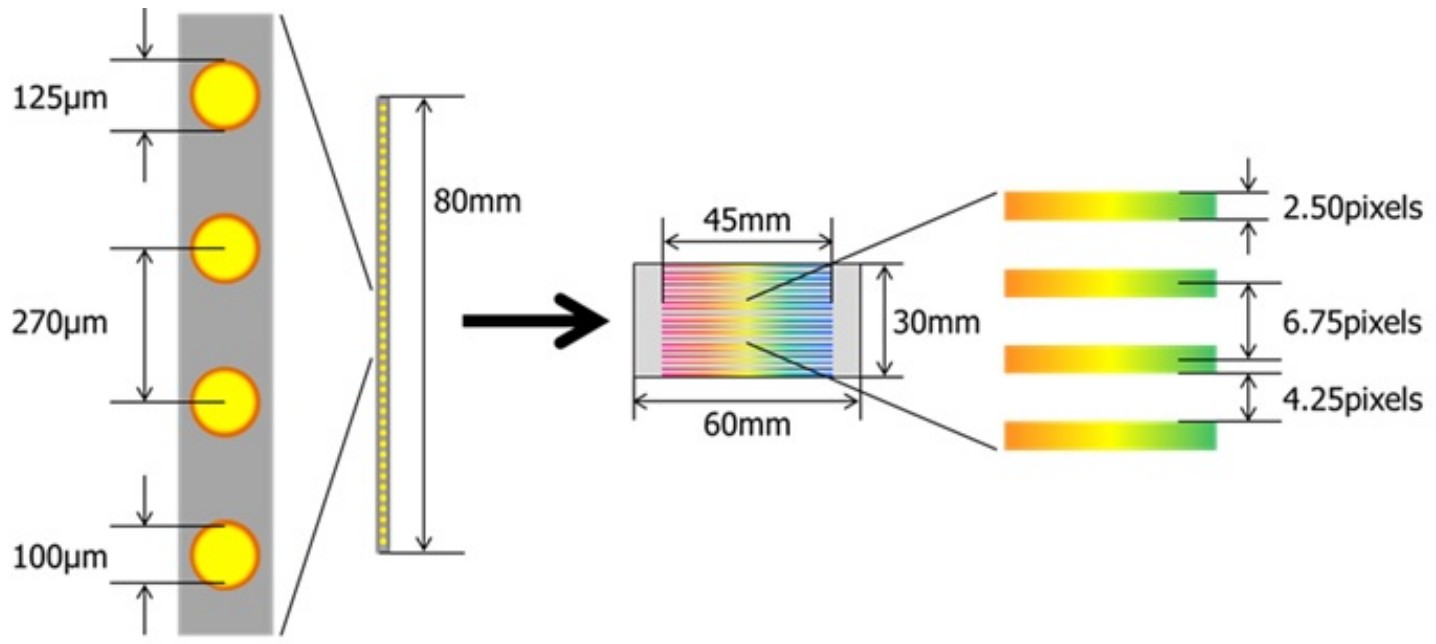

Figure 6: Mapping of the IFU slit-end to CCD.

# 3 DEVELOPMENT OF INTEGRAL FIELD UNITS

## 3.1 Mask from Photolithography

Photolithography as a chemical technique allows to precisely etch out parts from particular positions on a metal sheet. The position of the metal etching can be controlled to an accuracy of a few $\mu$m by performing photolithography accurately. One critical requirement to get properly shaped holes is the thickness of the metal foil which can be a maximum of 100 $\mu$m. Several masks are fabricated and then stacked to achieve a thickness of 1.5 mm. The mask making process has a yield of only 50% due to the chemical nature of the process. Thus we decided to produce twice the number of masks than what was actually required. Figure 8 illustrates photolithography through a schematic flowchart. The following points are to be taken care of to obtain masks which meet the requirements for an IFU.

1. Photolithography requires the purest form of metal foil. Impurities, if any, can act as a constructive or a destructive catalyst in the etching process leading to a higher or lower size of holes within the stipulated etching time. This reduces the yield of acceptable masks in a lot. The metal should be easily polishable, together with the fibers. The chosen thin foils have an extremely low percentage (<0.001%) of impurities. The glue, the metal, and the fiber have slightly different thermal expansion properties. The maximum difference in volume change between them is 0.1 $mm^3$ within the operating temperature (-5°to 25°) of the Devasthal observatory. Such volume change is <1% of the total volume of the fiber filled mask and hence acceptable.

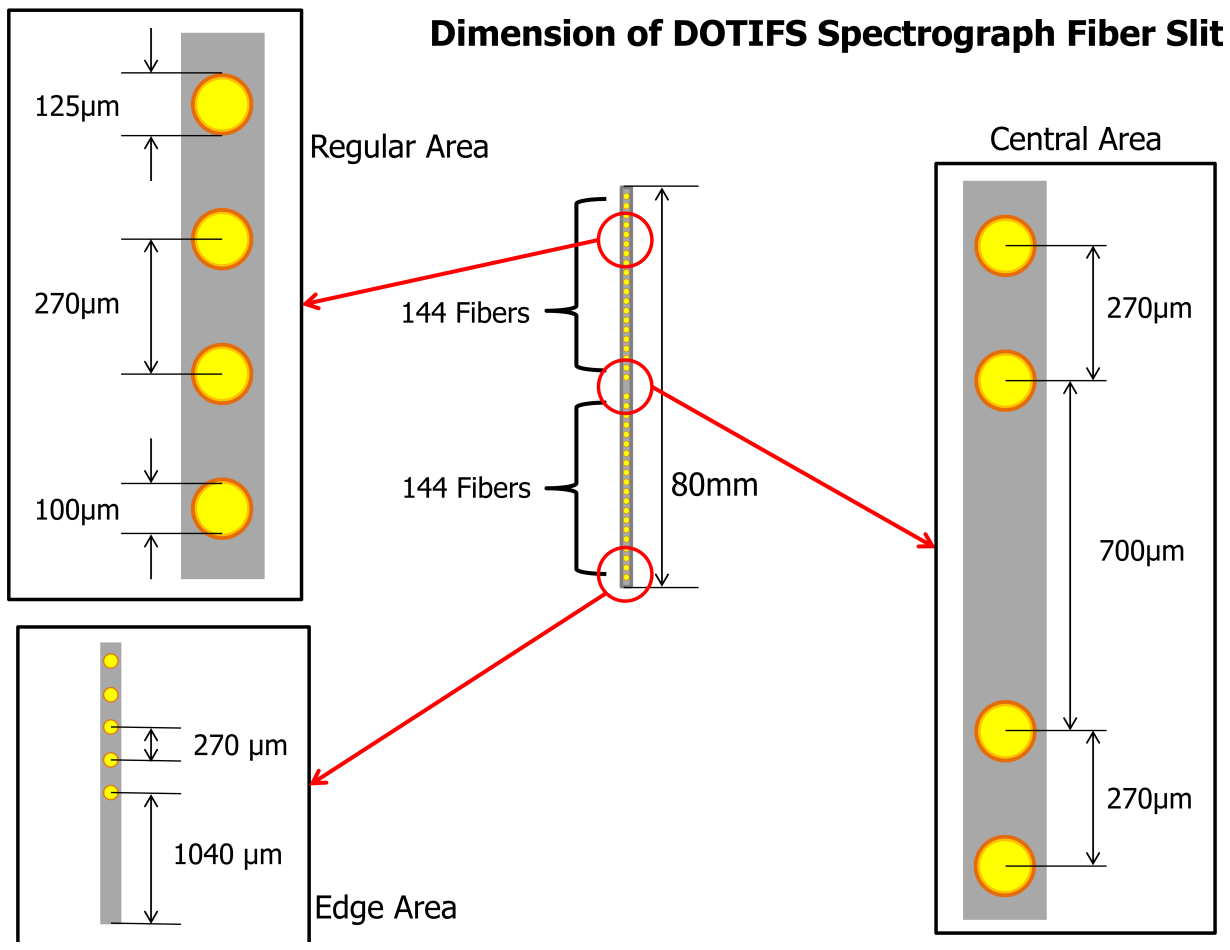

Figure 7: Dimensions of the IFU slit.

2. A metal foil (thickness 100 $\mu$m) of size 100×100 $mm^2$, is used for lithography. The two factors that determine the dimension of the foil are the number of masks that we want to produce at a time and the amount of the ultraviolet (UV) illuminable area. At least 25 masks are fabricated in a single process which is sufficient for achieving a stack thickness of 1.5 mm. We have used 12 masks in a stack, which corresponds to a total thickness of 1.2 mm. The remaining thickness of 300 $\mu$m is covered by layers of glue between each pair of foils which is of thickness 25 $\mu$m. The dimension of a mask is 3.6×3.2 $mm^2$. We have kept an additional 48 mm space on all four sides for ease of handling. On the other hand, the ultraviolet (UV) lamp assembly can illuminate an area corresponding to a circle of diameter 120 mm. The metal foil dimensions are large enough to create a 5 × 5 array of masks.

3. The foil is pre-processed before using it for photolithography to remove surface contaminants. It is cleaned with acetone and then placed between two thick steel blocks. The steel block surfaces are chrome plated to avoid introduction of any wiggle on the foil surface. The assembly is then heated and maintained at 250°C for three hours and then cooled down to room temperature. This process takes away any moisture and volatile contaminants present on the surface layer but creates a layer of metal oxide. We remove the oxide layer during the pre-processing stage by applying a dilute N/10 Hydrochloric acid.

4. A photoresist solution is prepared by mixing the dense photo-resist with the thinner solvent in a 7:3 volumetric ratio. The solution components are commercially available.

5. The pre-processed foil is then dipped entirely into the photoresist solution at a speed of 1 cm/s and pulled out at the same rate. This process is repeated three times, which leads to the deposition of a thin layer of photoresist on the metal foil. Typically, this layer is ∼10 $\mu$m thick. To achieve a thinner layer, the foil needs to be rotated at >1000 rotations per minute. Once a uniform layer has formed, the photoresist solution starts to harden due to evaporation of the solvent thinner. The hardening process usually takes about 30 minutes. Post baking may be used to achieve a uniform evaporation of solvent.

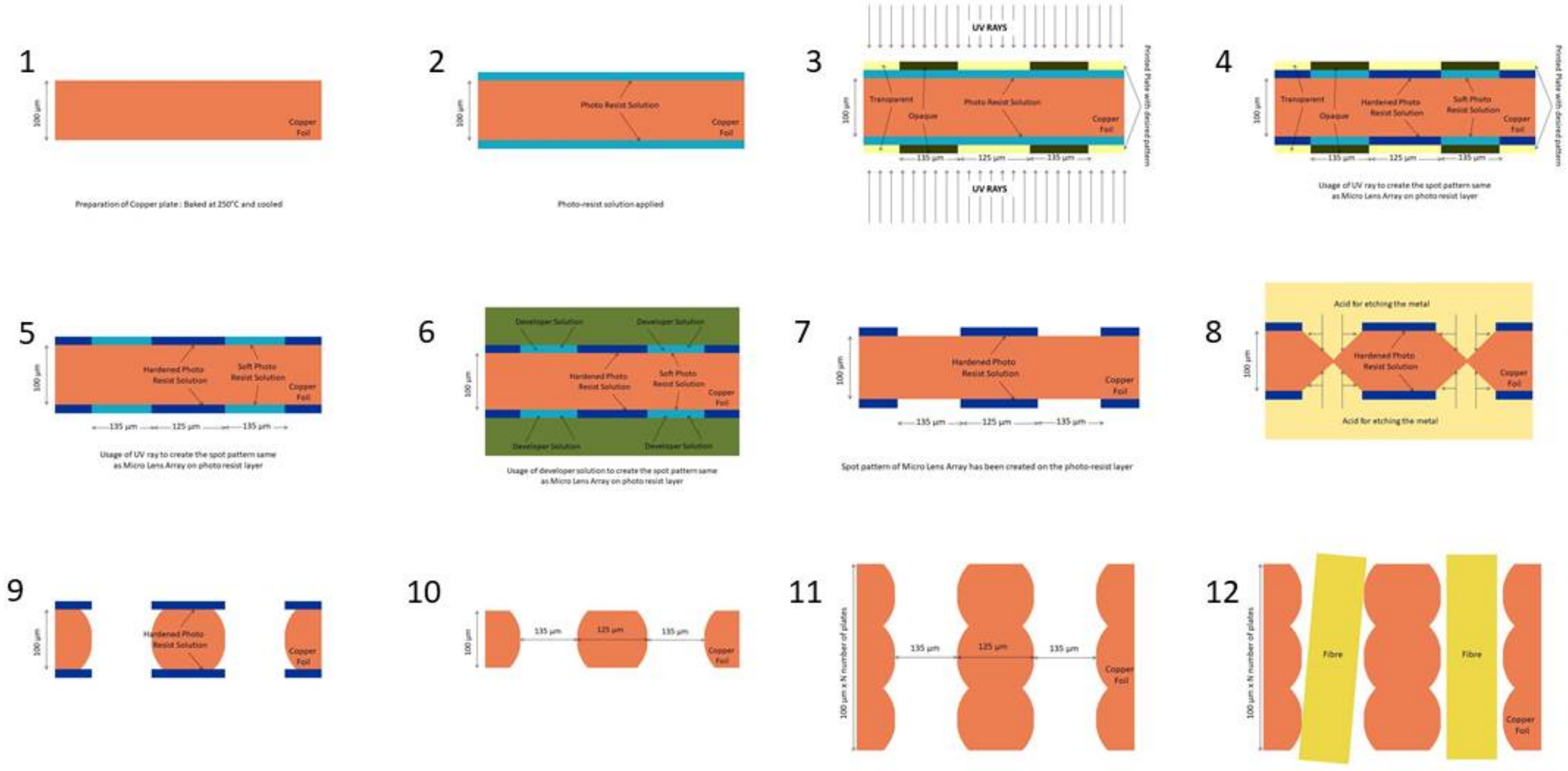

Figure 8: Schematic showing different steps of photolithography to obtain a mask.

6. In the next step, the desired pattern is created on a transparent sheet. The pattern consists of circular dark spots, which mimic the position and diameter of holes. The spot centroid pattern of each MLA is used to determine the location of the dark patches for the corresponding mask. The dark spots are opaque to the UV light. The remaining part of the transparent sheet does not hinder the transmission of the UV light. Etching performed on one side of the metal foil would result in a conical shape of the hole which is undesired as previously described. So etching is carried out on both sides of the foil simultaneously. The pattern and its mirror are printed on different transparent sheets. The two sheets are aligned face-to-face. The foil would be etched both along and across the surface. Hence the diameter of the spots is chosen to be 100 $\mu$m for a desired hole diameter of 130 $\mu$m.

7. The foil with the photoresist layer is then carefully placed between the two aligned sheets. Two glass plates are used to press the sheets tightly against the metal foil. The technique is called contact lithography. Then the glass-sheet-foil assembly is illuminated by a 125 W UV light source on both sides for 10 minutes. The UV light transmits through to the transparent areas of the sheet and chemically hardens the photoresist layer. However, the spots are opaque, and the photoresist layer beneath them remains chemically soft.

8. A developer solution is used to dissolve the chemically soft parts of the photoresist layer. The UV treated foil is dipped in the developer solution and stirred for a minute. The UV hardened areas of the photoresist layer do not get dissolved and stay intact. The foil is then dried, leading to evaporation of the developer solution. This process takes around 15 minutes.

9. The foil is then etched with Ferric Chloride solution (100 gm in 500 ml water) for 10 minutes. Ferric Chloride etches out the exposed metal areas from both sides to create the desired hole array. The foil is then cleaned with running water to remove traces of the etching reagent.

10. At this point, the foil still has the hardened photoresist coating. Tri-chloroethylene is used to dissolve the remaining photoresist layer before cleaning the foil with N/10 Hydrochloric acid.

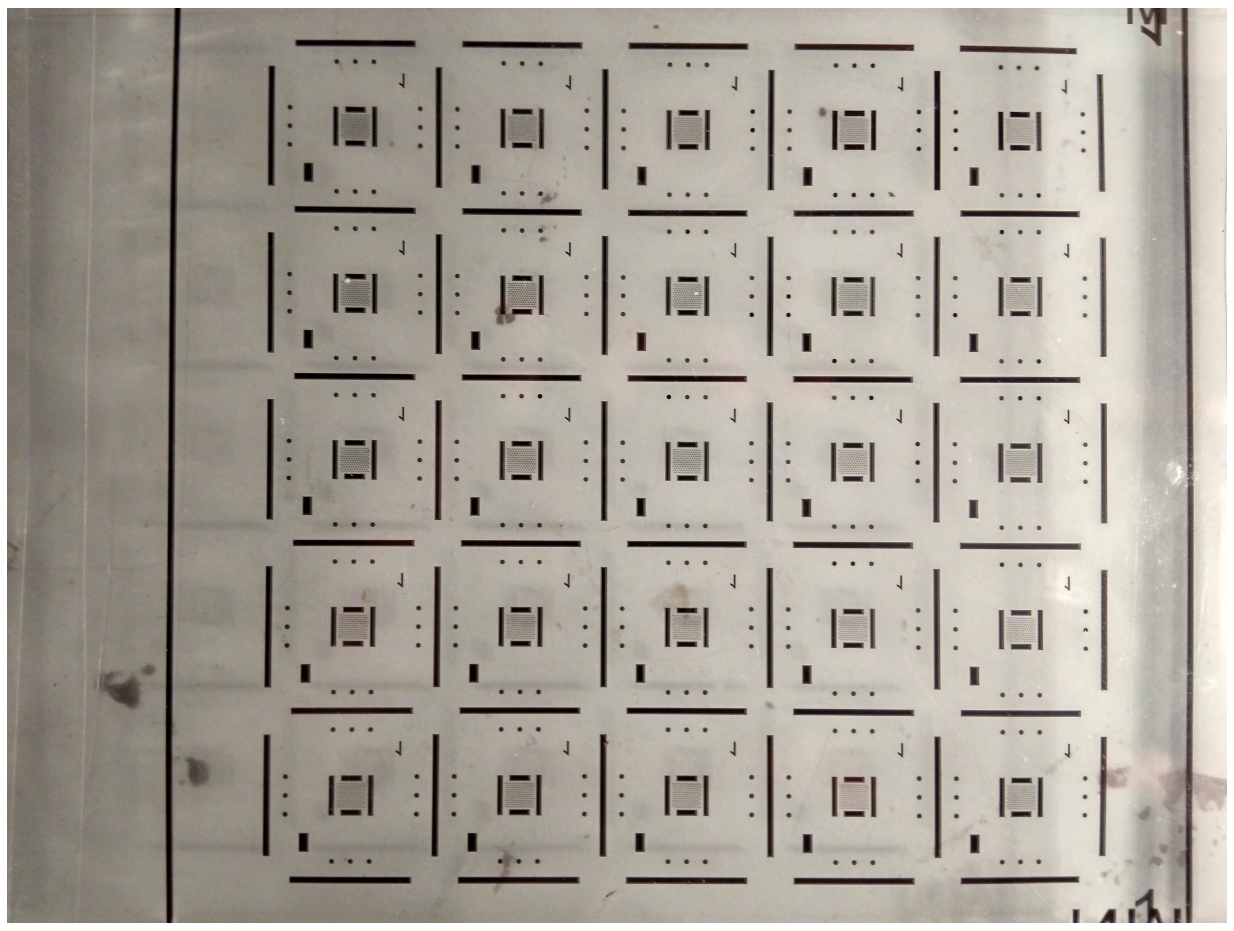

Figure 9: Image of a 2 layer well aligned transparent sheet with 25 patterns.

### 3.2 Mask Selection

Once photolithography is performed, the foil would generate 25 masks from the two well aligned transparent sheets with 25 patterns, as shown in Figure 9. These masks are characterized before use. The centroid positions and diameters of the 144 holes are measured for each mask by using back-illuminated images (e.g., Figure 11). The masks are divided into four categories, 1, 2, 3 and 4. Every hole in a category 1 masks have diameters which are between 125-130 $\mu$m and is best suitable for our use. Category 2 and 3 have hole diameters ranging from 125-135 $\mu$m and 125-140 $\mu$m respectively. These masks can be used at the back of the mask stack as shown in Figure 10. Category 4 masks have holes with diameter >140 $\mu$m and thus are not suitable for our use. It is empirically found that category 1, 2, 3 and 4 have 10%, 25%, 20% and 45% yield respectively. Our intention is to use at least four category one masks for each IFU which required typically two lithography foils to fabricate a single IFU.

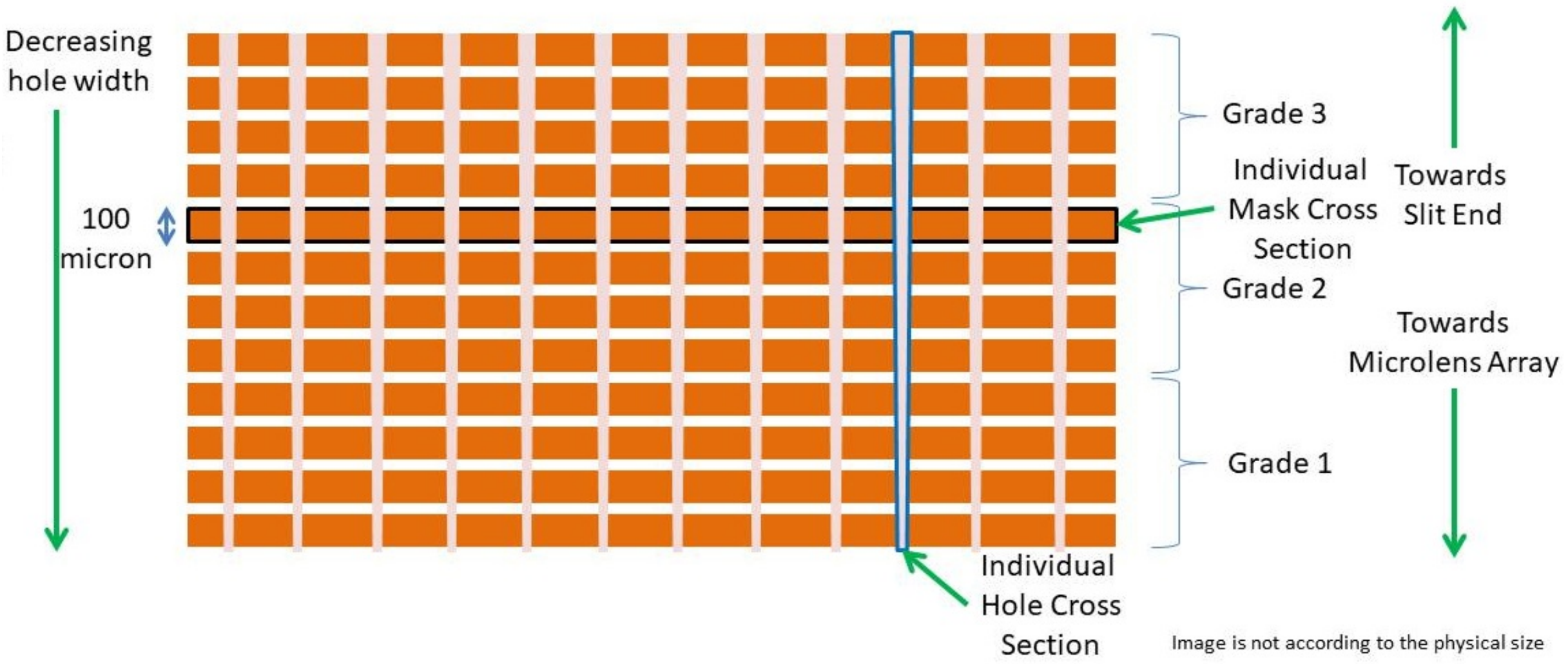

Figure 10: Schematic cross section of the mask stack describing the use of different category masks in different parts of the mask stack.

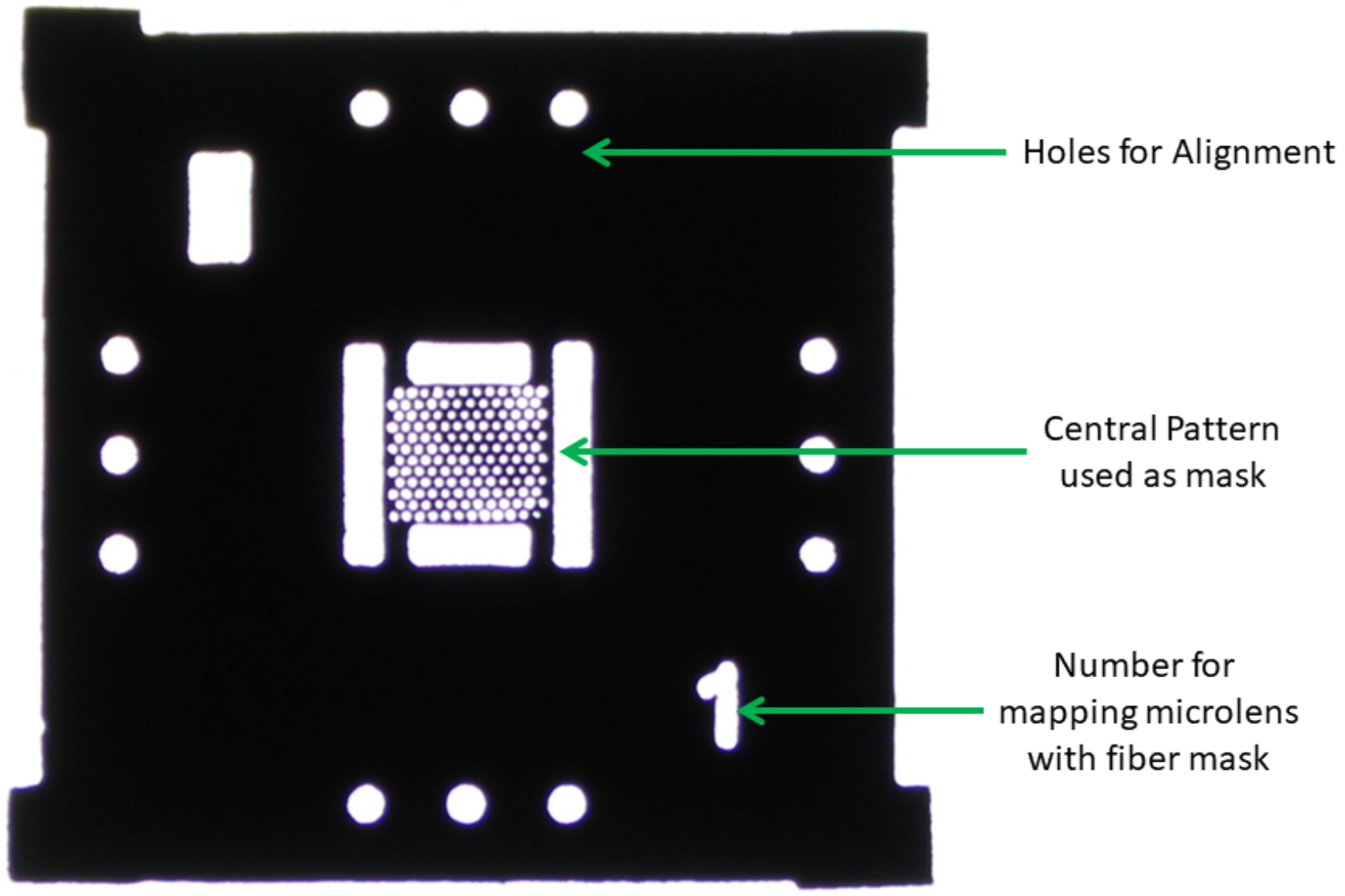

Figure 11: Back-illuminated image of a single mask ready to be used for stacking. The entire mask including the central pattern is created using lithography.

## 3.3 Mask Alignment & Fiber Insertion

The chosen twelve masks are divided into three groups to ensure that different categories of masks remain separate from each other. Each group contains four masks of the same category. These groups of masks are then held by bench vices. The fibers are inserted in the masks after dissolving the jacket of the initial 20 mm length with hot sulphuric acid. Remaining length of the fiber was kept with jacket. Removal of the jacket reduces the fiber diameter to 110 $\mu$m, which in turn makes the insertion easier. Once all the 144 fibers are inserted, the masks are brought together and glued. The process of gluing should avoid trapping of any air bubbles between the fibers. We have used the Epotek 301-2 (volumetric ratio of 2:1 between hardener and softener) for the gluing purpose. After gluing, the fiber mask assembly is kept inside a heat bath. The heat bath is made of a single 250 W lamp encased inside an airtight metal container. The glue takes around 24 hours to get cured. The first prototype IFU has a missing fiber (refer to Figure 14). Collapsing the masks together after insertion caused the fibers to come off, as the fiber without the jacket did not fit tightly into the hole. It was not possible to reinsert the fiber as the array is too congested with fibers protruding from both sides of the mask stack.

## 3.4 Polishing of Fiber Bundle & Mask Stack

The fiber bundle is required to be polished to a surface finish of $\sim$0.3 $\mu$m. The cured glue, the three-layer fiber and the mask material (copper) have different coefficients of hardness. If hard abrasive agents are used the rate of removal of the material will be different for different surfaces. So we have chosen Aluminum Oxide powder as a relatively softer abrasive material. We have used standard commercially available Aluminum Oxide of grain sizes of 13.5, 9, 3, 1 and 0.3 $\mu$m. The powders are used in the order of decreasing size to achieve a smoother polishing in a gradual manner. A fiber polishing apparatus was built previously at IUCAA, which is shown in Figure 13. It consists of a circular disk (of diameter 500 mm) rotated by a 24 V DC motor. A polishing cloth is glued to the disk, and a slurry of Aluminum Oxide powder with water is prepared on top of the fabric. The fiber bundles were polished for 12-14 hours with the slurry of each grain size and checked under a microscope. An unpolished fiber will show scratch marks on its tip and produces spots on the detector with lower intensity compared to completely polished fibers as shown in Figure 12. Figure 14 shows the back-illuminated view of the focal-plane-end of an IFU after polishing. The dark spot shows a missing fiber.

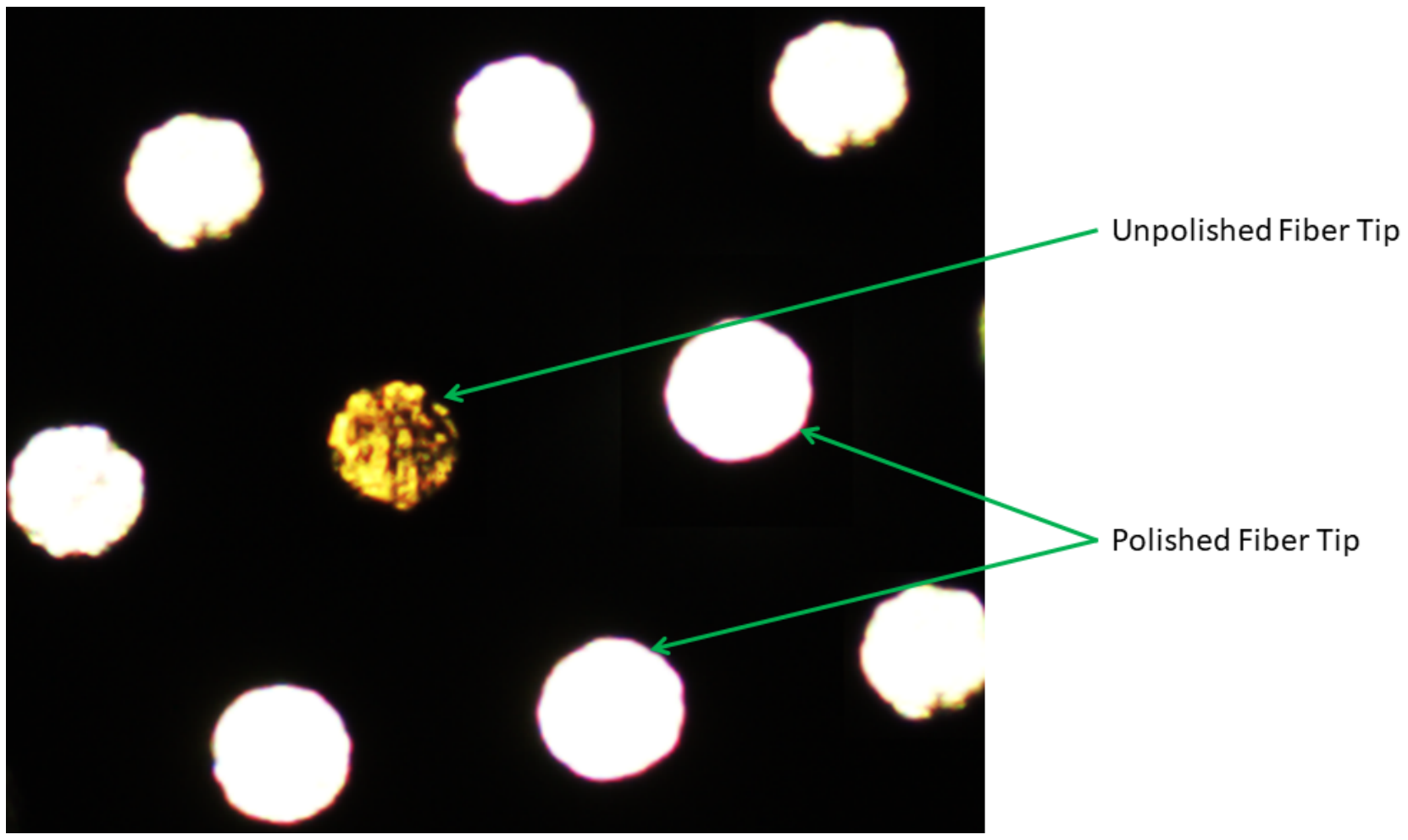

Figure 12: Photograph demonstrating the difference between polished and unpolished fibers.

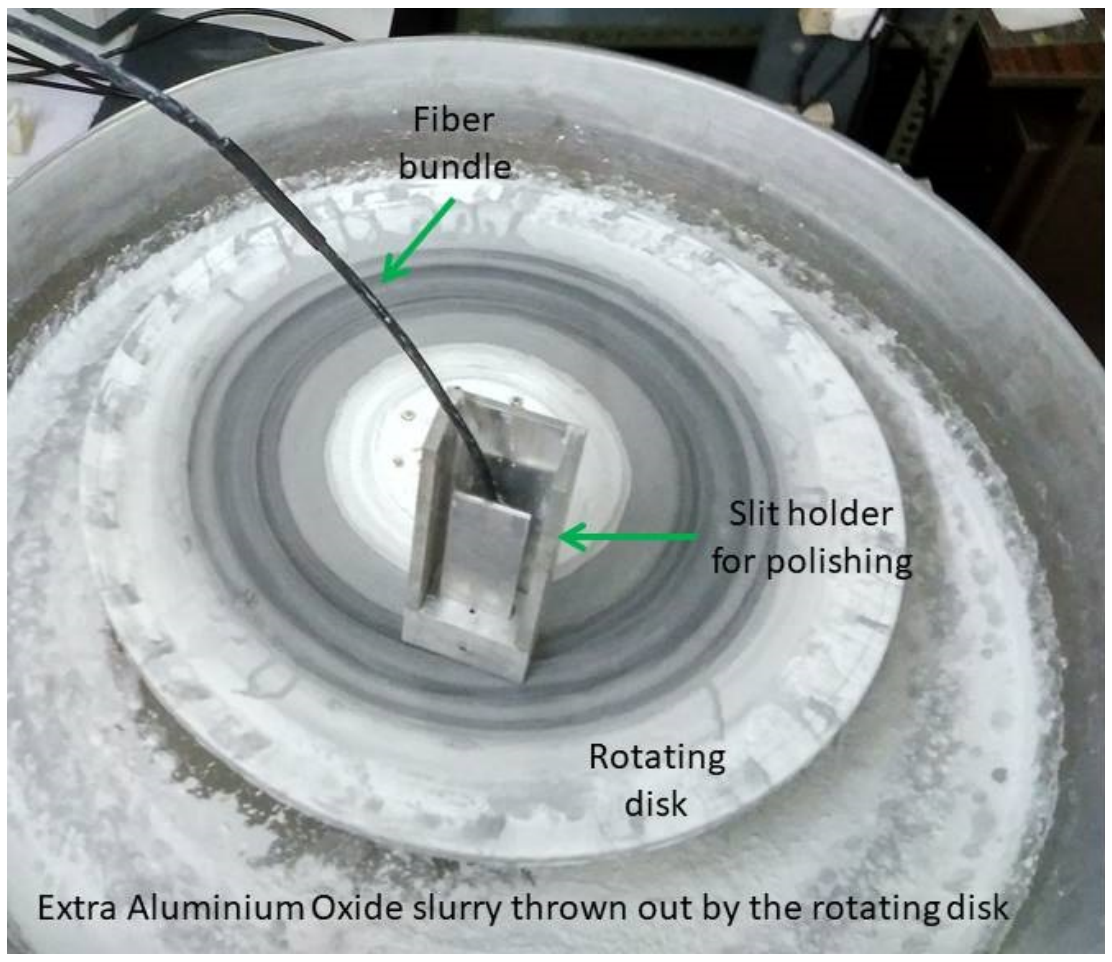

Figure 13: Photograph of the polishing setup to polish the slit-end of the IFU. The setup consists of a DC motor which rotates the circular disk.

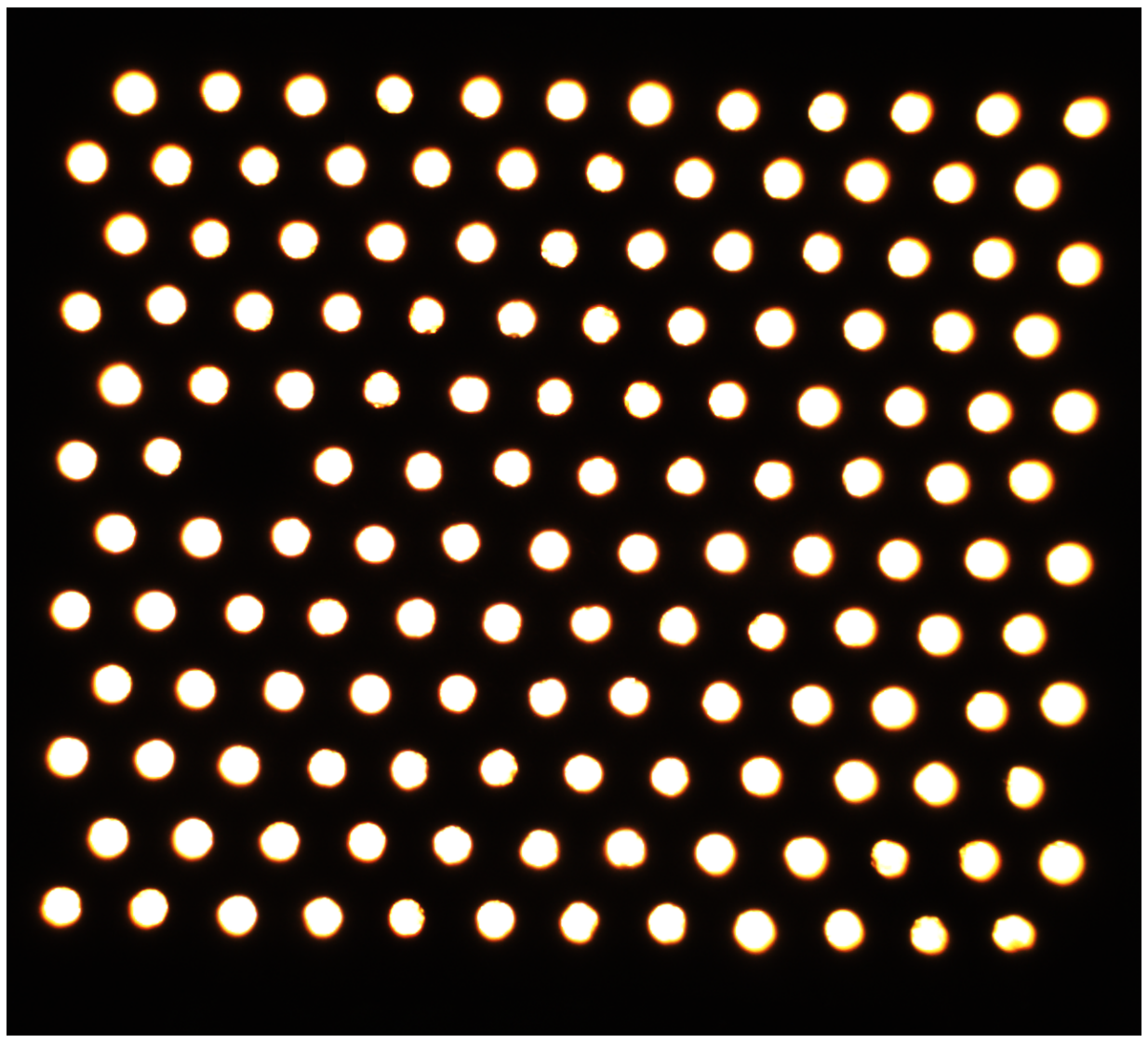

Figure 14: Back illuminated image of the fiber filled mask after gluing and polishing. Dark spot shows the single dropped fiber. This prototype IFU has a missing fiber. Collapsing the masks together after insertion causes the fibers to come off, as the fiber without the jacket did not fit tightly into the hole. It was not possible to reinsert the fiber as the array is too congested with fibers protruding from both sides of the mask stack.

### 3.5 Development of Slit-end

Figure 15 shows the mechanical design of the 'U' shaped grooves for holding the fibers. Two such blocks hold the fibers from the two IFUs and sit together in a mechanical holder attached to the spectrograph base plate, as shown in Figure 16. We have made the U-groove block from Aluminum with standard wire cutting technology using a 100 $\mu$m wire. Figure 17 shows an edge-on view of the U-grooves. Another piece of Aluminum is used to cover the U-groove. The assembly creates holes of semi-circular cross section. We have used an epoxy glue for gluing fibers to the U-grooves. The mapping pattern of fibers from the focal-plane-end to the slit-end is shown in Figure 18.

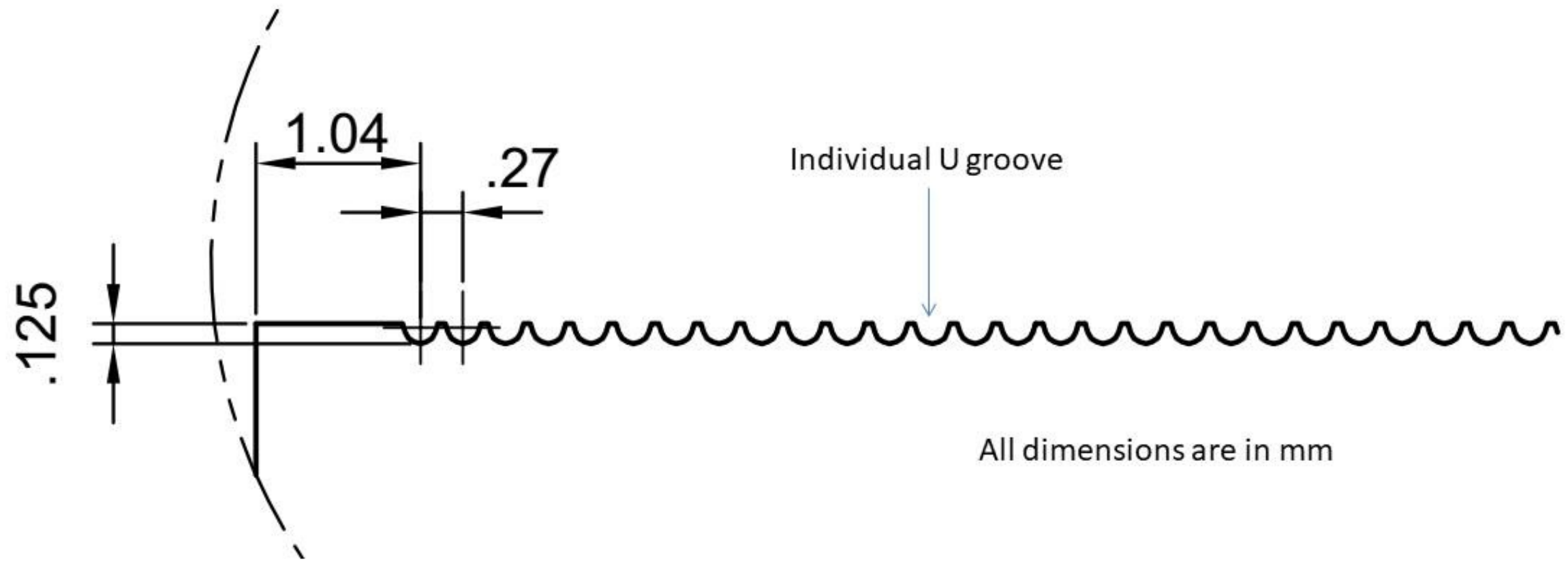

Figure 15: Schematic showing dimensions of the IFU slit.

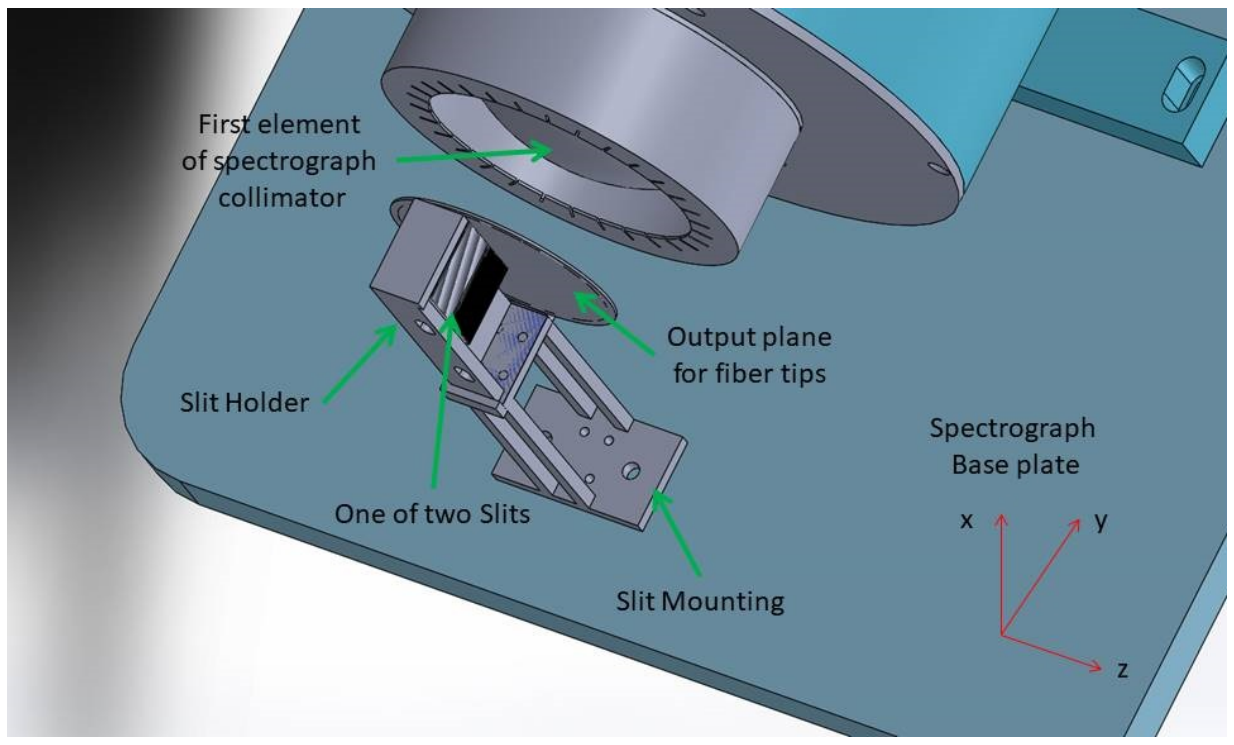

Figure 16: Solidworks snapshot shows the scheme for mounting the slits at the input of the DOTIFS spectrograph.

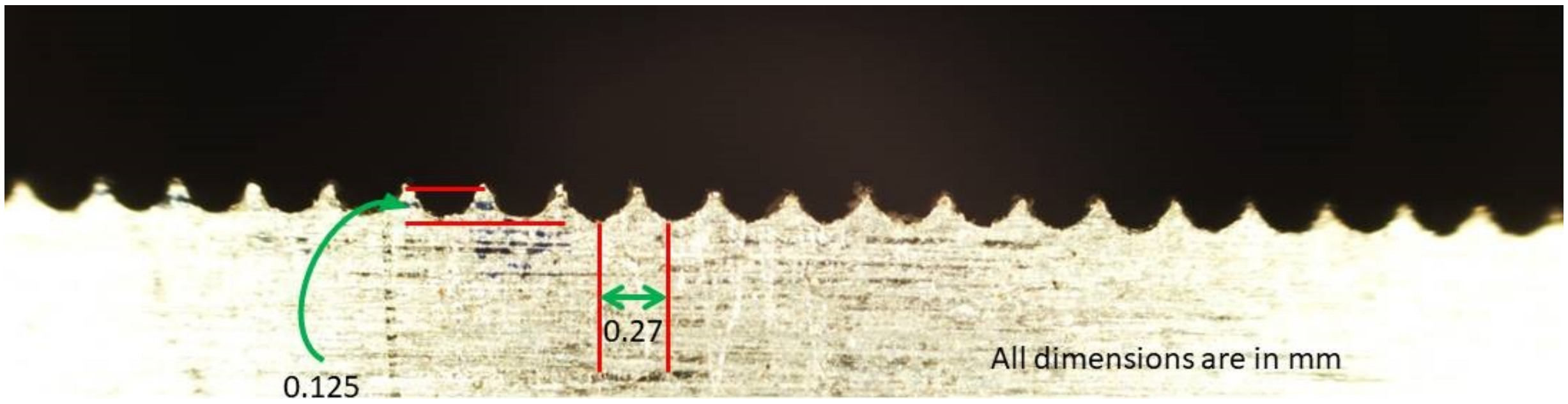

Figure 17: Microscopic edge-on view of the slit manufactured by wire cutting method.

Each fiber was inserted in a slit array without glue during its insertion in the mask stack. Once all the 144 fibers are inserted, the fibers were transferred to a slit with glue (in liquid phase) maintaining their relative positions. The assembly is then kept to cure in a heat chamber as described in subsection 3.3. We have used the existing fiber polishing apparatus as described in subsection 3.4 and as shown in Figure 13 to polish the slit after chopping off the protruding fiber lengths. Figure 19 shows the image of the slit-end after polishing.

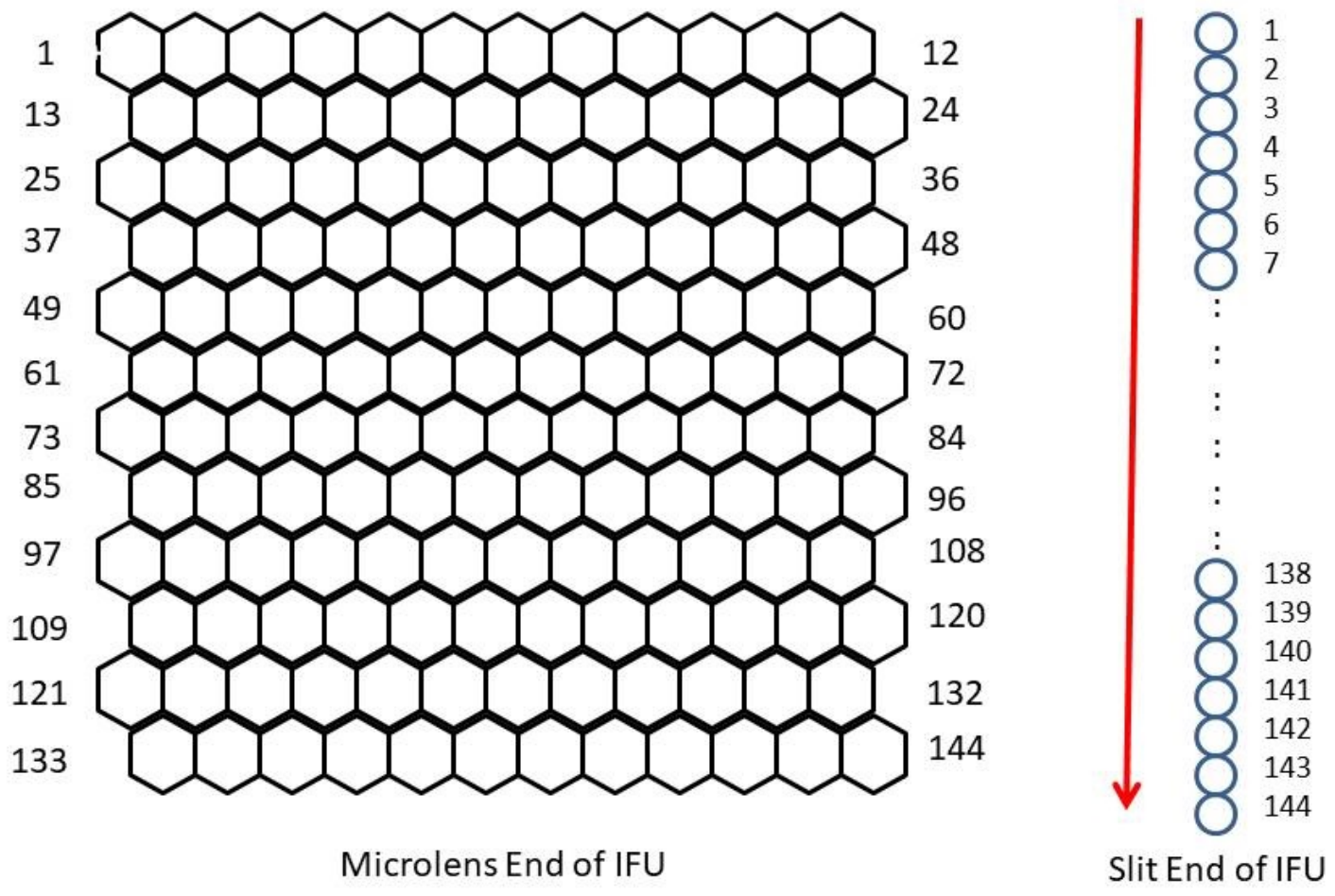

Figure 18: Schematic showing the mapping of fibers from the focal-plane-end to the slit-end.

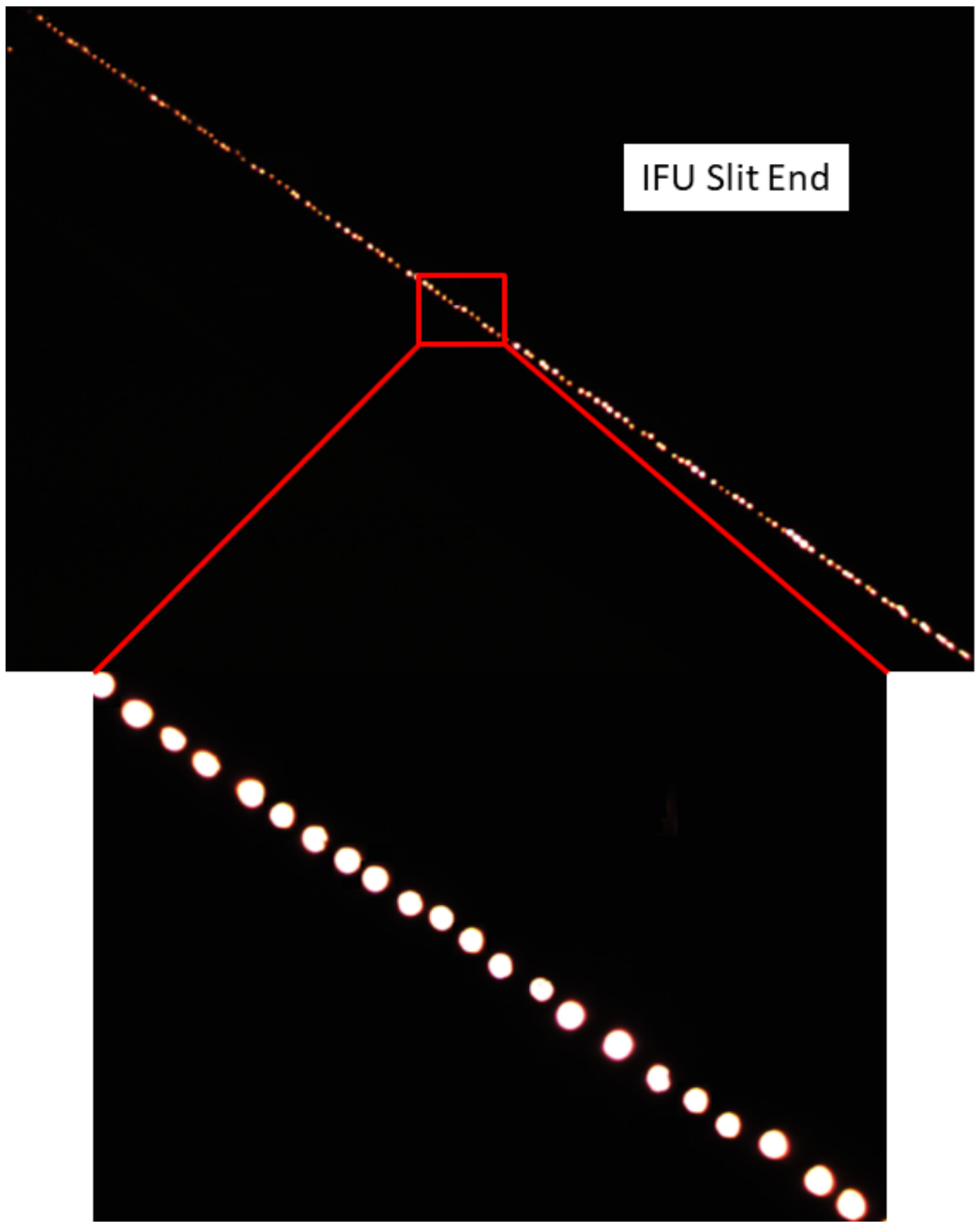

Figure 19: Photograph of the IFU slit-end in back-illuminated condition.

### 3.6 Characterization of the assembly

The polished assembly is characterized in terms of the positions of the fibers and their relative throughput. For this, back-illuminated images of the focal-plane-end and the slit-end of the IFU are obtained. An automated script is developed to find the centroid of each fiber spot and then it is compared with the spot centroid pattern created by the MLA. Figure 20 shows the error associated with the positioning of each fiber. It is found that the maximum positioning error for any fiber is ∼4.9 $\mu$m. Thus all the fibers are well within the position error budget provided by the design. This avoids light loss due to fiber positioning. A similar centroid position measurement is performed for the slit against the ideally expected positions of the fibers, and the result of the same is shown in Figure 21. The maximum error of fiber positioning in the slit is 24 $\mu$m which corresponds to 0.6 pixels, compared to the gap of 6.75 pixels between two spectra.

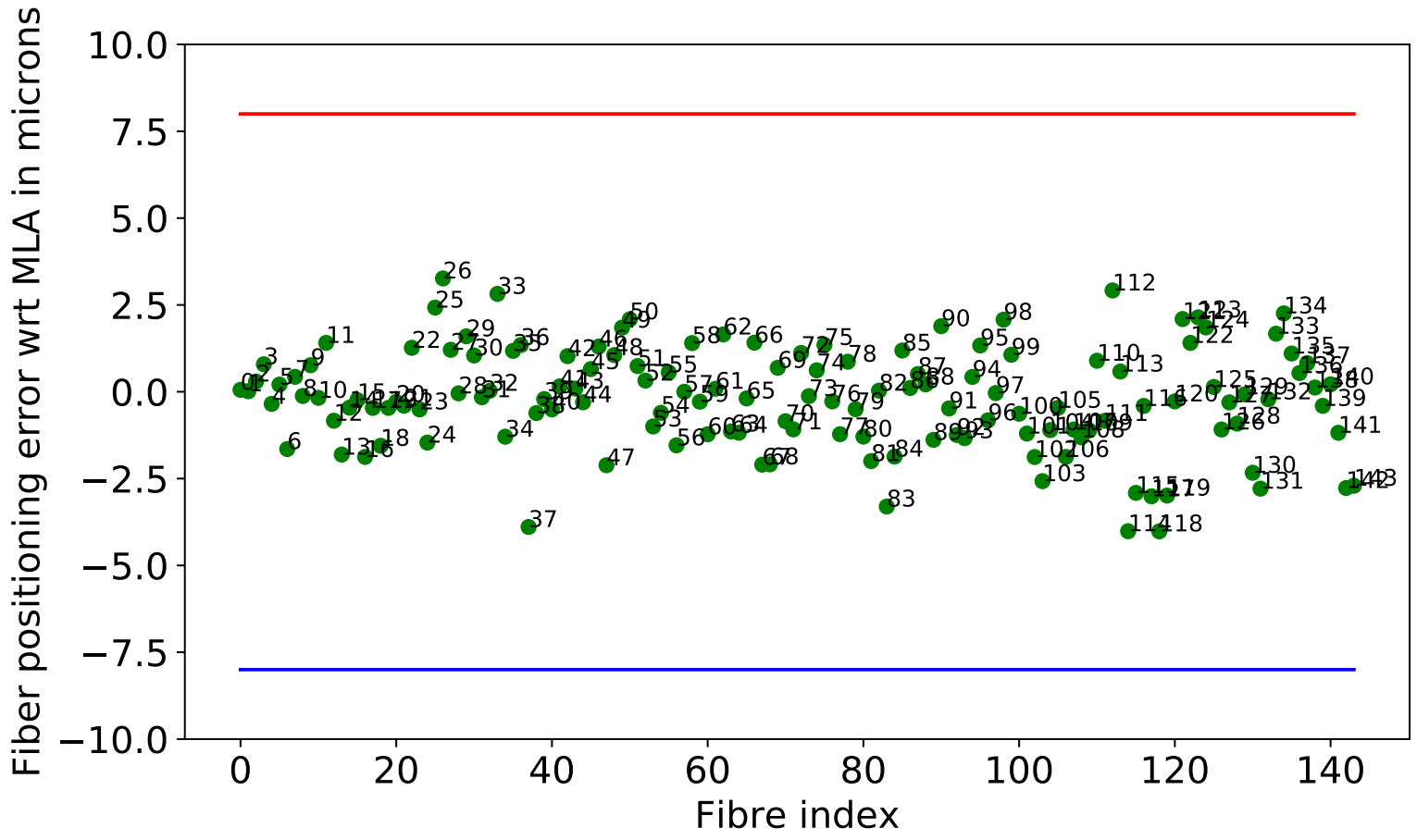

Figure 20: Plot of the positioning error of fibers using the mask with the respective microlens. The red and the blue line shows the boundaries of maximum allowed positive and negative errors respectively. The annotated numbers denote the fiber positions as shown in figure 18

### 3.7 Alignment of Fiber Mask and Microlens Array

The fiber filled mask and the MLA are aligned before gluing them. An aligned microlens with a back illuminated fiber-mask casts an image on the detector with appropriate imaging optics. A Zemax simulation predicts that for an ideal alignment (with imaging arrangement as shown in Figure 22), the diameter of this image should be 175 $\mu$m. The slit-end is illuminated by a collimated light beam. The light is then carried by the fiber to the flat surface of the microlens.The output of the microlens is reimaged on to an ST9 camera. The diameter of the image will change drastically as shown in Figure 23 if the alignment is not perfect.

This setup was not possible to achieve in the laboratory as the biconvex lens predicted by Zemax has different curvatures on the two sides of the lens while the one we used has a symmetrical shape. We found the diameter to be 200 $\mu$m during the alignment in the laboratory (Figure 24). Replacing the ideal lens by the real lens in Zemax simulation has indeed produced a spot of 200 $\mu$m diameter. The image of the pattern is analyzed by measuring the relative throughput of all fibers. The throughput is measured by comparing the total pixel counts within a circular aperture of 10 pixels diameter (1 pixel corresponds to 20 $\mu$m physical size). Once we confirmed that the MLA and the mask were aligned and the relative throughput difference between fibers is less than 7% (as shown in Figure 25), we applied the Norland Optical Adhesive 68 glue. A 125 watt UV light is used for 15 minutes, for curing process. Once cured, the assembly is kept in the heat bath for 12 hours. The heating process increases the chemical bond formation rate between the MLA and the fiber mask surface with the glue.

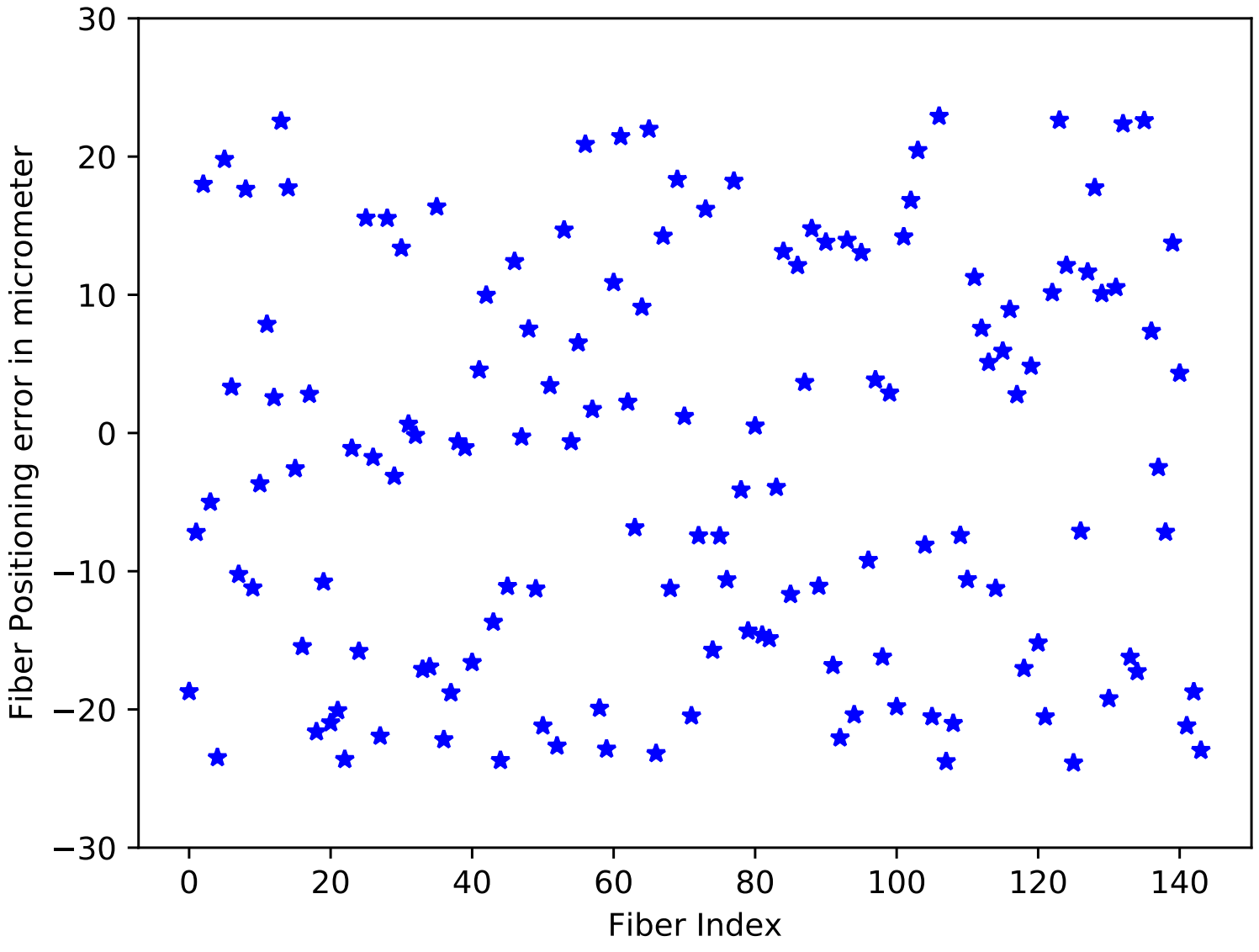

Figure 21: Plot of the positioning errors of the fibers on the slit, with respect to the ideal position.

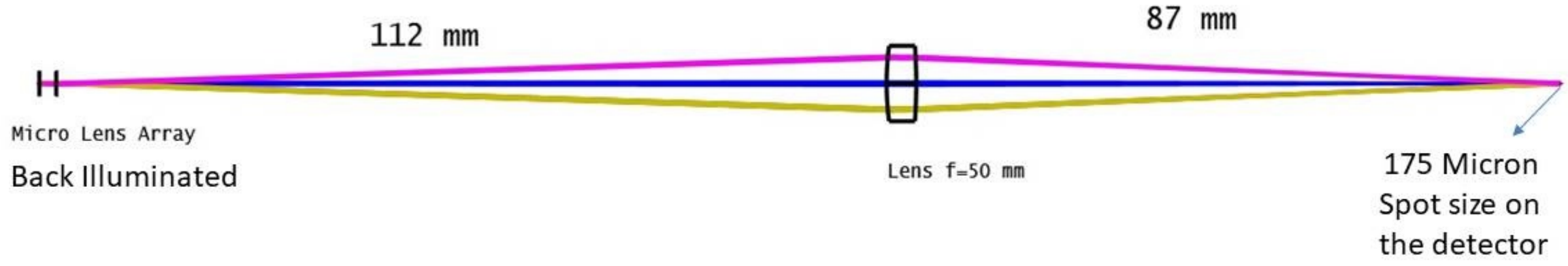

Figure 22: Schematic of the microlens fiber alignment system from zemax simulation. The fiber bundle needs to be back illuminated.

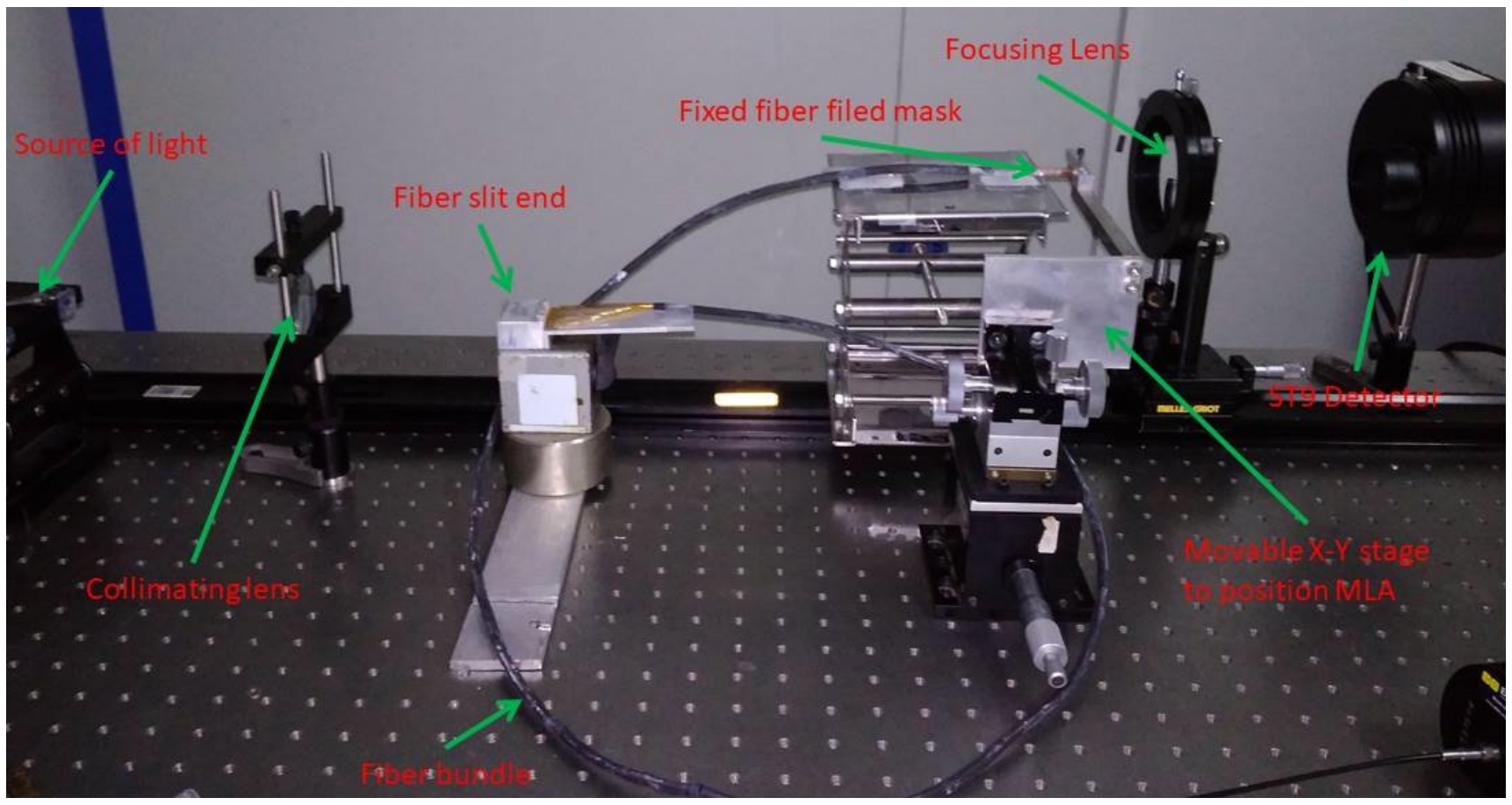

Figure 24: Photograph showing the microlens fiber alignment scheme. The fiber bundle is back illuminated.

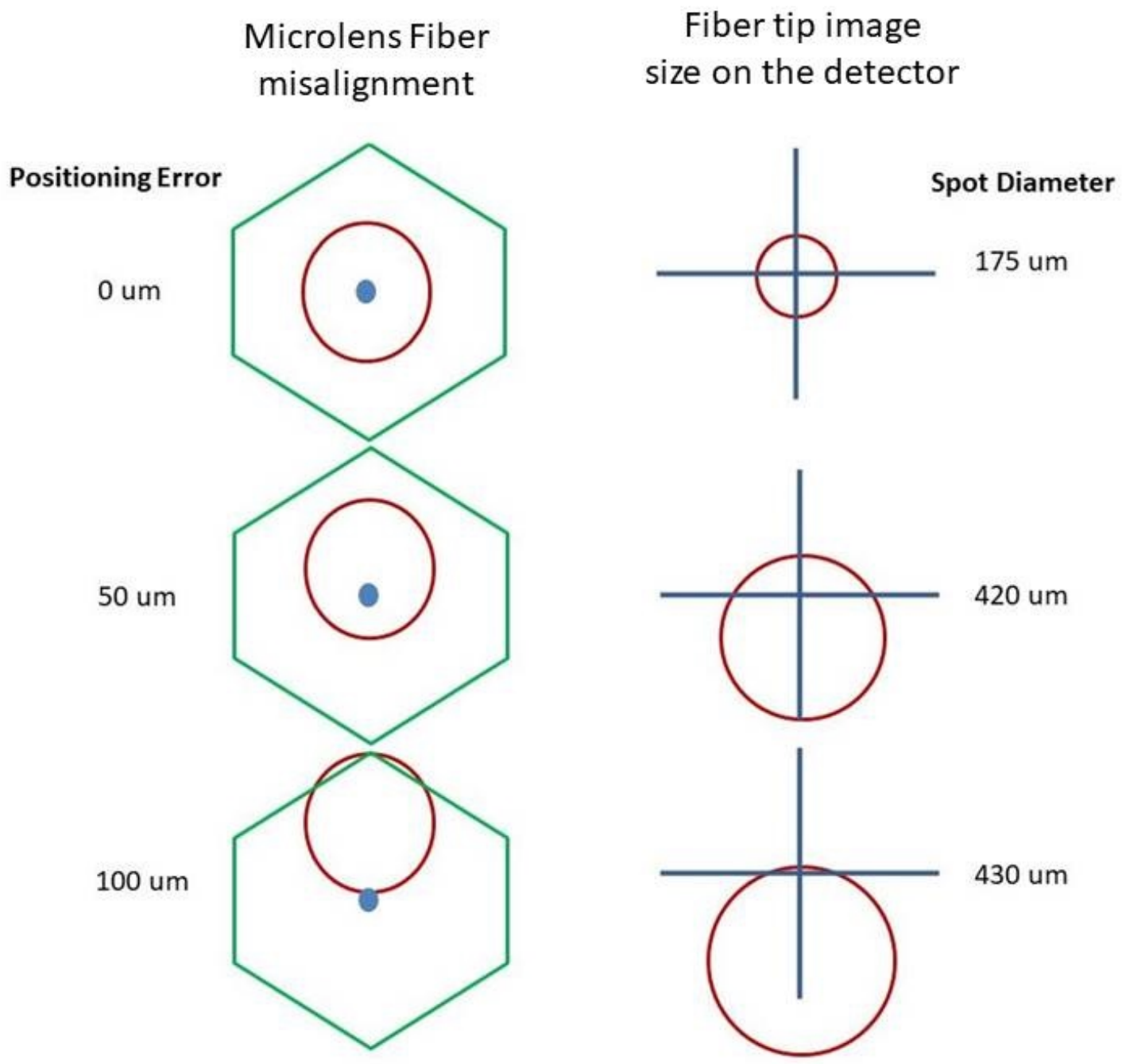

Figure 23: Schematic showing the change in size of the fiber tip for various positions of the fiber with respect to the microlens.

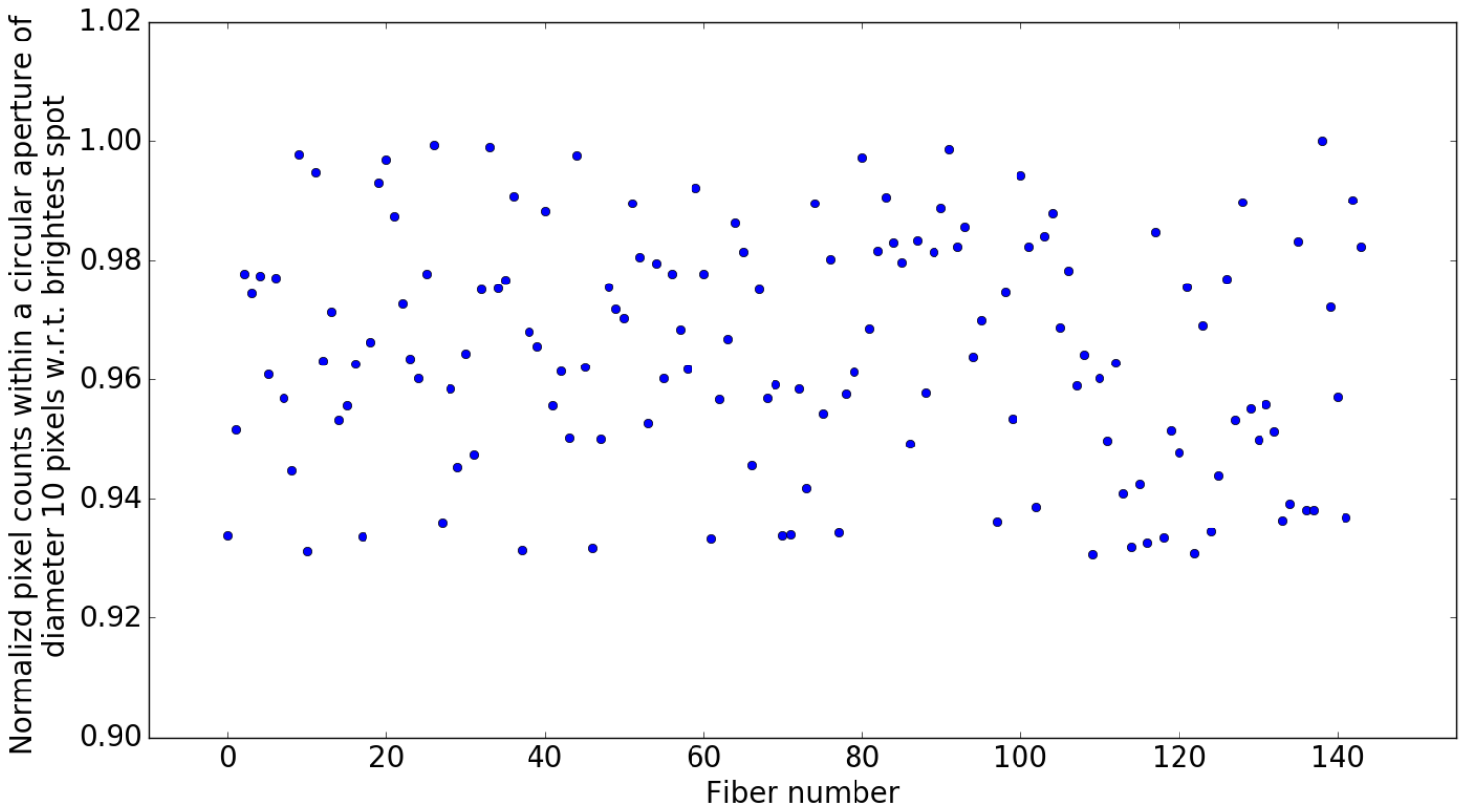

Figure 25: Plot showing the relative difference in the fiber intensities after polishing.

# 4 SUMMARY

An end-to-end integral field unit is developed using the technique of photolithography. The IFU consists of fibers positioned behind a microlens array by means of a mask. The microlens array is customized based on the optical design. Each array is characterized in terms of the individual lens positions within an uncertainty of ± 4 $\mu$m . The position of the lenslets is transfered on thin (100 $\mu$m) copper foils by photolithography. The lithographed foils are stacked to form masks with a positional accuracy of better than 4.9$\mu$m for the holes. The slit-end of the IFU is made by wire cut U-grooves. The maximum positioning error corresponds to 0.6 pixels at the CCD, compared to a gap of 6.75 pixels between two consecutive spectra. The mapping of fibers from the focal-plane-end to the slit-end is ensured during the insertion stage. Finally, the microlens array and the corresponding fiber filled masks are aligned and glued. The fiber-to-fiber throughput variation is only 7% (peak to peak) with respect to the highest throughput fiber.